\documentclass[12pt]{article}
\pdfoutput=1

\usepackage[bulletsep]{collref}
\usepackage{amssymb,graphicx}
\usepackage[intlimits]{amsmath}
\usepackage{bbm}
\usepackage[small]{subfigure}

\usepackage{MnSymbol}

\makeatletter \@addtoreset{equation}{section} \makeatother

\makeatletter
\let\old@startsection=\@startsection
\let\oldl@section=\l@section
\renewcommand{\@startsection}[6]{\old@startsection{#1}{#2}{#3}{#4}{#5}{#6\mathversion{bold}}}
\renewcommand{\l@section}[2]{\oldl@section{\mathversion{bold}#1}{#2}}
\makeatother

\makeatletter
\let\old@makecaption=\@makecaption
\def\@makecaption{\small\old@makecaption}
\makeatother

\renewcommand{\leq}{\leqslant}
\renewcommand{\geq}{\geqslant}

\def\KK{{\cal K}}

\begin{document}


\begin{flushright}\footnotesize
\texttt{NORDITA-2014-119} \\
\texttt{UUITP-13/14}
\vspace{0.6cm}
\end{flushright}

\renewcommand{\thefootnote}{\fnsymbol{footnote}}
\setcounter{footnote}{0}

\begin{center}
{\Large\textbf{\mathversion{bold} Strong-Coupling Phases 
\\  of Planar $N=2^*$ Super-Yang-Mills Theory}
\par}

\vspace{0.8cm}

\textrm{
K.~Zarembo}
\vspace{4mm}

\textit{Nordita, KTH Royal Institute of Technology and Stockholm University,
Roslagstullsbacken 23, SE-106 91 Stockholm, Sweden}\\
\textit{Department of Physics and Astronomy, Uppsala University\\
SE-751 08 Uppsala, Sweden}\\
\textit{Institute of Theoretical and Experimental Physics\\ B. Cheremushkinskaya 25, 117218 Moscow, Russia}\\
\vspace{0.2cm}
\texttt{zarembo@nordita.org}

\vspace{3mm}


\par\vspace{1cm}

\textbf{Abstract} \vspace{3mm}

\begin{minipage}{13cm}
The $\mathcal{N}=2^*$ theory (mass deformation of $\mathcal{N}=4$ Super-Yang-Mills) undergoes an infinite number of quantum phase transitions in the large-$N$ limit. The phase structure and critical behavior can be analyzed with the help of supersymmetric localization, which reduces the problem to an effective matrix model.  We study its strong-coupling phase.
\end{minipage}

\end{center}

\par\vspace{0.3cm}
\begin{flushright}{\it To Andrei Alexeevich Slavnov on occasion of his 75th birthday}
\end{flushright}



\renewcommand{\thefootnote}{\arabic{footnote}}
\setcounter{footnote}{0}

\section{Introduction}

The $\mathcal{N}=4$ superconformal Yang-Mills  (SYM) theory possesses remarkable hidden symmetries and its strong-coupling behavior can be understood in detail by virtue of the AdS/CFT duality. Among massive theories,
the ones  most likely to inherit this intrinsic simplicity are
relevant perturbations of $\mathcal{N}=4$ SYM. We concentrate the mass perturbation of $\mathcal{N}=4$ SYM  known as the $\mathcal{N}=2^*$ theory. While this theory has less symmetries and more complicated dynamics compared to $\mathcal{N}=4$ SYM, some of the non-perturbative results do extend from $\mathcal{N}=4$ to $\mathcal{N}=2^*$. One such result is the exact computation of the path integral on $S^4$ \cite{Pestun:2007rz} by supersymmetric localization \cite{Witten:1988ze}. 

The path integral on $S^4$ reduces to a zero-dimensional matrix model in any  $\mathcal{N}=2$ gauge theory. In the $\mathcal{N}=4$ case the matrix model is Gaussian   \cite{Erickson:2000af,Drukker:2000rr,Pestun:2007rz},  while in the $\mathcal{N}=2^*$ case it is not  \cite{Pestun:2007rz}, moreover the localization matrix model of the $\mathcal{N}=2^*$ theory has a very complicated  phase structure at large-$N$ (the regime where the gauge/string duality operates in the simplest way) \cite{Russo:2013qaa,Russo:2013kea}. The model  has been studied in detail in some corners of the parameter space \cite{Buchel:2013id,Russo:2013kea,Chen:2014vka}. In particular,  its asymptotic strong-coupling solution is known  and allows for non-trivial tests of holographic duality beyond AdS/CFT \cite{Buchel:2013id}. In particular, one can calculate the free energy and the Wilson loop expectation values for sufficiently large contours directly from path integral and then compare them with the geometric data of the  supergravity dual in flat space  \cite{Pilch:2000ue}, as well as on $S^4$ \cite{Bobev:2013cja}.

Compactification on $S^4$ may be viewed as formal means to pick a unique vacuum and to make the path integral well-defined without imposing boundary conditions (this point of view was articulated in \cite{Russo:2013sba}). The radius $R$ of the four-sphere then plays the r\^ole of an IR regulator, to be sent to infinity at the end of the calculation.  The resulting theory in infinite volume turns out to possess a rather intricate phase structure, undergoing an infinite number of quantum phase transitions when the 't~Hooft coupling is varied from zero to infinity \cite{Russo:2013qaa}. The large-$N$ phase transitions are  common in matrix models \cite{Gross:1980he,Wadia:2012fr}, but those in the $\mathcal{N}=2^*$ theory are of a novel type \cite{Russo:2013kea}, and the associated critical behavior has not been completely understood. 

The exact solution of the localization matrix model in the weak-coupling phase was obtained in \cite{Russo:2013qaa}, following the methods developed in \cite{Hoppe,Kazakov:1998ji}, while the strong-coupling phases have been only studied in the infinitesimal vicinity of the first phase transition \cite{Russo:2013kea}. The goal of this paper is to investigate the strong-coupling phases of $\mathcal{N}=2^*$ SYM in more detail. 

A phase structure similar to that of $\mathcal{N}=2^*$ SYM was found in three-dimensional massive theories amenable to localization \cite{Kapustin:2009kz}. The associated matrix model is quite a bit simpler and can be solved analytically in the decompactification limit \cite{Anderson:2014hxa}. We are going to transplant the method of   \cite{Anderson:2014hxa} to four dimensions. Although not leading to a full analytic solution, this method provides a detailed picture of the vacuum structure across the whole phase diagram.

\section{Localization matrix model}

The field content of $\mathcal{N}=2^*$ SYM 
consists of the vector multiplet, containing the gauge fields $A_\mu $, scalars $\Phi $, $\Phi '$ plus fermions, and two CPT conjugate hypermultiplets  that contain complex scalars $Z$, $\tilde{Z}$ and fermions. All fields are in the adjoint representation of the $SU(N)$ gauge group. An overall energy scale is set by the explicit mass term for the hypermultiplets. The mass parameter in the Lagrangian will be denoted by $M$. Once $M$ is set to zero, the Lagrangian becomes that of the  $\mathcal{N}=4$ SYM. 

The gauge symmetry of the $\mathcal{N}=2^*$ theory is spontaneously broken by an expectation value of  the vector multiplet scalar:
\begin{equation}\label{vac}
 \left\langle \Phi \right\rangle=\mathop{\mathrm{diag}}\left(a_1,\ldots ,a_N\right),
\end{equation}
which Higgses $SU(N)$ down to $U(1)^{N-1}$ and gives masses to almost all of the fields, except for the gauge bosons of the unbroken $U(1)^{N-1}$. The masses of the $ij$ components of the vector multiplet ($i$ and $j$ are the color indices) are $|a_i-a_j|$, while the hypermultiplet masses are $|a_i-a_j\pm M|$.  The eigenvalues $a_i$ and $a_j$ which are distance $M$ apart form a resonance -- there is a massless hypermultiplet associated with them. The phase transitions in the matrix model are caused precisely by these resonances.

The path integral of the $\mathcal{N}=2^*$ theory on $S^4$ localizes to an eigenvalue integral over the Coulomb moduli $a_i$  \cite{Pestun:2007rz}:
\begin{equation}
 Z=\int d^{N-1}a\,
\prod _{i<j}\frac{(a_i-a_j)^2H^2(a_i-a_j)}{H(a_i-a_j-M)H(a_i-a_j+M)}
\,{\rm e}\,^{-\frac{8\pi^2N}{\lambda}\sum_i a_i^2 },
\end{equation}
where $\lambda =g_{\rm YM}^2N$ is the 't~Hooft coupling.
The Gaussian term in the action descends from the gravitational coupling of the vector multiplet to the background metric of $S^4$. The rest of the degrees of freedom have been integrated out leaving behind the one-loop contribution that depends on the single function
\begin{equation}\label{functionH}
 H(x)\equiv \prod_{n=1}^\infty \left(1+\frac{x^2}{n^2}\right)^n \,{\rm e}\,^{-\frac{x^2}{n}} .
\end{equation}
The exact partition function also contains an instanton contribution which however remains exponentially suppressed at large $N$ \cite{Russo:2013kea}. As we are interested in the large-$N$ limit, we set instantons to zero from the outset. 

Localization cannot compute arbitrary correlation functions. Among the few that can be calculated is the Wilson loop for the big circle of $S^4$, whose expectation value maps to the exponential average in the matrix model \cite{Pestun:2007rz}:
\begin{equation}\label{Wilson}
 W(C_{\rm ircle})\equiv \left\langle \frac{1}{N}\,\mathop{\mathrm{tr}}{\rm P}\exp
 \oint_{C_{\rm ircle}}ds\,\left(iA_\mu \dot{x}^\mu +\Phi |\dot{x}|\right)
 \right\rangle \stackrel{{\rm loc.}}{=}\left\langle \frac{1}{N}\sum_{i}^{}\,{\rm e}\,^{2\pi a_i}\right\rangle.
\end{equation}
Another quantity of interest is the vacuum susceptibility, which can be related to the mean square of the matrix eigenvalues:
\begin{equation}\label{susceptibility}
 \chi \equiv \frac{\lambda ^2}{8\pi ^2N^2}\,\,\frac{\partial \ln Z}{\partial \lambda }=\left\langle \frac{1}{N}\sum_{i}^{}a_i^2\right\rangle.
\end{equation}
This again can be computed exactly using localization.

In the leading order of the large-$N$ expansion the eigenvalue integral can be evaluated in  the saddle-point approximation. The saddle-point equations, upon introducing the eigenvalue density
\begin{equation}\label{eigenvalue-density}
  \rho (x)=\left\langle \frac{1}{N}\sum_{i=1}^{N}\delta \left(x-a_i\right)\right\rangle,
\end{equation}
take the form of a singular integral equation:
\begin{equation}
 \label{nnstar}
\strokedint_{-\mu}^\mu dy\, \rho(y)
\left(\frac{1}{x-y} -\KK(x-y)+\frac{1}{2}\,\KK(x-y+M)+\frac{1}{2}\,\KK(x-y-M)\right)= \frac{8\pi^2}{\lambda}\ x,
\end{equation}
where $\KK=-H'/H$. This equation was studied in a number of papers \cite{Russo:2012kj,Buchel:2013id,Russo:2013qaa,Russo:2013kea,Russo:2013sba,Chen:2014vka}, but its general analytic solution is not known\footnote{An equation with a similar kernel arises in the $\mathcal{N}=4$ SYM compactified on the squashed four-sphere \cite{Huang:2014pda,Crossley:2014oea}. In that case the argument in the kernel is shifted by the squashing parameter of the sphere rather than by the mass of the hypermultiplet. Once this model is mass-deformed it also develops a non-trivial phase structure \cite{Marmiroli:2014ssa}.}.

The equations above are written in the dimensionless variables with radius of the four-sphere set to one. The dependence on $R$ is recovered by rescaling $a_i\rightarrow a_iR$, $M\rightarrow MR$, $x\rightarrow xR$ and $y\rightarrow yR$, such that the mass and the Coulomb moduli get back their normal scaling dimensions. In this paper we regard $R$ as a regularization parameter, and are mainly interested in the decompactification limit  $R\rightarrow \infty $. The saddle-point equation then considerably simplifies. The argument of the  kernel function $\KK(Rx)$ becomes large, and the function can be replaced by its asymptotics at infinity: $\KK(x)\rightarrow x\ln x^2$. Differentiating the resulting equations twice, we arrive at 
\begin{equation}\label{intnormal1}
 \strokedint_{-\mu }^\mu
 dy\,\rho (y)  \left( \frac{2}{x-y} - \frac{1}{x-y+M}-\frac{1}{x-y-M} \right) =0.
\end{equation}
This is the saddle-point equation of the decompactified theory, and it is this equation that we are going to study in the rest of the paper.  

This equation looks very simple, but
in spite of its simple appearance, it describes rather complicated behavior, including  an infinite number of phase transitions which happen as the ratio $\mu /M$, the only dimensionless parameter in the game, changes from zero to infinity. The equation can be solved exactly for sufficiently small $\mu /M$   \cite{Russo:2013qaa,Russo:2013kea}, using the methods of \cite{Hoppe,Kazakov:1998ji}. This analytic solution describes the weak-coupling phase of the theory and terminates  at  a fourth-order quantum phase transition at $\mu =M/2$  \cite{Russo:2013qaa,Russo:2013kea}. We are going to concentrate on the strong-coupling phases at $\mu >M/2$, where no analytic results were available so far.

In the decompactification limit, the density blows up at the edges of the interval:
\begin{equation}\label{boun}
 \rho (x)\sim \frac{\,{\rm const}\,}{\sqrt{\mu \mp x}}\qquad \left(x\rightarrow \pm \mu \right).
\end{equation}
These boundary conditions are slightly unfamiliar. In matrix models, the square-root asymptotics is typically found \cite{Brezin:1977sv}.  In fact, the exact solution  in finite volume has this normal boundary behavior. But when $R$ (more precisely, $MR$) becomes large, the density develops two peaks near the endpoints of the eigenvalue distribution, whose width is proportional to $1/R$ and the hight grows with $R$. In the strict $R\rightarrow \infty $ limit the structure of the peaks is no longer resolved, and the peaks become infinite one-over-square-root spikes, hence leading to the boundary conditions (\ref{boun}). The qualitative reasons for this behavior are explained in  \cite{Russo:2012ay}. 

Another way to see that the square root boundary conditions are inconsistent with eq.~(\ref{intnormal1}) is to notice that the solution with such boundary conditions would have to be unique, while the unconventional boundary conditions (\ref{boun}) admit a one-parameter family of solutions \cite{Gakhov}. This extra degree of freedom is built in the equations, because they are invariant under rescaling of the density by an arbitrary constant: $\rho (x)\rightarrow C\rho (x)$. The rescaling ambiguity is fixed by normalizing the density to one. In a more familiar setting, the normalization condition constitutes an extra constraint that determines the endpoint of the eigenvalue distribution.

Now imposing the normalization condition is not sufficient to find $\mu $. To fix $\mu $, we need to go one step back, and write down an equation that is obtained from (\ref{nnstar}) by single differentiation. This will be an integral form of (\ref{intnormal1}), with a specific constant of integration. It thus suffices to impose this condition at one single point, which we choose to be $x=-\mu $:
\begin{equation}\label{formintegrated}
 \int_{-\mu }^{\mu }dy\,\rho (y)\ln\frac{\left|M^2-(\mu +y)^2\right|}{(\mu +y)^2}=\frac{8\pi ^2}{\lambda }\,.
\end{equation}
The correctly normalized solution of the saddle-point equations is unique for any
given $\mu $. If we manage to find this solution, the last equation determines $\lambda $ as a function of $\mu $ and $M$. Instead of inverting this function, it is more convenient to  treat $M$ and $\mu $ (rather than $M$ and $\lambda $) as independent variables.  

Knowing $\mu $ we can compute the Wilson loop expectation value from (\ref{Wilson}). In the decompactification limit $R\rightarrow \infty $, the Wilson loop gets the largest contribution from the largest eigenvalues, and thus obeys the perimeter law:
\begin{equation}\label{Perimeter}
 \ln W(C_{\rm ircle})=2\pi R\mu .
\end{equation}
The perimeter law  should apply to any sufficiently large contour, not just to the big circle of $S^4$. This has been checked holographically and the coefficient of proportionality was found to match with the one found from solving the matrix model at strong coupling \cite{Buchel:2013id}.

The qualitative structure of the solution to (\ref{intnormal1}) depends on the relationship between the width of the eigenvalue distribution $2\mu $ and the mass $M$. If $M>2\mu $, only the first term in the kernel has a pole in the integration region, the equation is of the Hilbert type and is solvable in elliptic integrals \cite{Russo:2013qaa}. As mentioned above, the solution hits a singularity at $\mu =M/2$, equivalently at   $\lambda _c\simeq 35.425$, signalling a transition to a new phase \cite{Russo:2013qaa}. The origin of the phase transition is easy to understand. When $2\mu $ exceeds $M$, some pairs of eigenvalues become separated by $M$. The resonant pairs correspond to massless hypermultiplets (actually, nearly massless, with $1/N$ accuracy). These resonances lead to poles in the two other terms in the integral equation which were regular in the weak-coupling phase. As a result,  the density develops two cusps at $\pm (M-\mu )$,  the resonance images of the endpoint positions. 

The structure of the resonances in the infinitesimal vicinity of the phase transition was delineated in \cite{Russo:2013kea}. It was found that  the cusps have a lambda-like shape, such that the density approaches a finite limiting value from the inside of the cusp and has an inverse-square-root singularity on the outside. Thermodynamic singularity associated with the appearance of the first pair of resonances is likely to be a fourth-order phase transition \cite{Russo:2013kea}. The resonances multiply with the increase of the coupling, appearing in pairs each time $2\mu $ crossed an integer multiple of $M$. The system thus undergoes an infinite number of  phase transitions on the way from weak to strong coupling. At strictly infinite coupling, an analytic solution with infinitely many cusps was obtained in \cite{Chen:2014vka}, by first solving the system at  finite $MR$ and then taking the decompactification limit $R\rightarrow \infty $. Since the corner $(\lambda ,MR)=(\infty ,\infty )$ of the phase diagram is an accumulation point of an infinite number of phase transitions, the physics may as well depend on the direction along which one approaches the critical point. We show, nevertheless, that the solution of  \cite{Chen:2014vka} also emerges if one takes the limit $R\rightarrow \infty $ first and then approaches the critical point by taking $\lambda \rightarrow \infty $.

 We will study the strong-coupling phases of $\mathcal{N}=2^*$ SYM by combining the ansatz  of \cite{Russo:2013kea} with the method that has led to an exact solution of the mass-deformed ABJM theory \cite{Anderson:2014hxa}. We start with the simplest strong-coupling phase in which the density has two resonances.

\section{Two-resonance phase}

When $\mu $ is very small ($\mu \ll M$), the system is in the weak-coupling phase, and the density is equal to $1/\pi \sqrt{\mu ^2-x^2}$. As $\mu $ grows and becomes comparable to $M$ the density gets deformed but qualitatively does not change too much. The exact shape is actually known for any $\mu <M/2$  \cite{Russo:2013qaa}, even though the analytic expression is not particularly simple. Once $\mu $ exceeds $M/2$, the density changes dramatically by
developing resonance cusps at $\pm\mu \mp nM$.
For the time being we assume that $2\mu >M$ but  $\mu <M$, such that there are exactly two resonances at $M-\mu $ and $-M+\mu $. 

We then take the following ansatz\footnote{This ansatz was initially suggested to us by D.~Volin.}, that generalizes the structure found   at $2\mu -M\ll M$:
\begin{equation}\label{rho-r}
 C\,\rho (x)=
\begin{cases}
 r(x), & {\rm }x\in\left(-M+\mu ,M+\mu \right)
\\
 \frac{4}{3}\,r(x)+\frac{2}{3}\,r(x-M), & {\rm }x\in\left(M-\mu ,\mu \right)
 \\
  \frac{4}{3}\,r(x)+\frac{2}{3}\,r(x+M), & {\rm }x\in\left(-\mu ,-M+\mu \right).
\end{cases}
\end{equation}
The function $r(x)$ can be pictured as an analytic continuation of the density from the region between the two resonance points. The key assumption is that such analytic continuation is well defined on the whole interval $(-\mu ,\mu )$ and obeys the same boundary conditions (\ref{boun}) as the density, in other words has inverse square root branch points at $x=\pm \mu $. 
The shift operator in the map from
$r(x)$ to $\rho (x)$ takes into account points connected to $x$ by massless hypermultiplets (notice that for $x$ between $-\mu +M$ and $\mu -M$ such resonance points lie outside the eigenvalue distribution). It will become clear shortly why the coefficients of the shift operator are taken to be $4/3$ and $2/3$. An overall normalization factor $C$ is introduced for later convenience, and is eventually fixed by the normalization condition.

Quite importantly, the map from $r(x)$ to $\rho (x)$ is non-local. As a  consequence, the endpoint singularities of $r(x)$ induce discontinuities in the middle of the eigenvalue interval. Even if $r(x)$ itself is a smooth function between $-\mu $ and $\mu $ (which we tacitly assume), the density $\rho (x)$ will have induced singularities at $M-\mu $ and $-M+\mu $. These singularities arise as images of the endpoints under the map $x\rightarrow x\pm M$, and are associated with resonances on the nearly massless hypermultiplets.

\begin{figure}[t]
\begin{center}
 \subfigure[]{
   \includegraphics[height=4cm, trim=0cm 0cm 0cm 0cm] {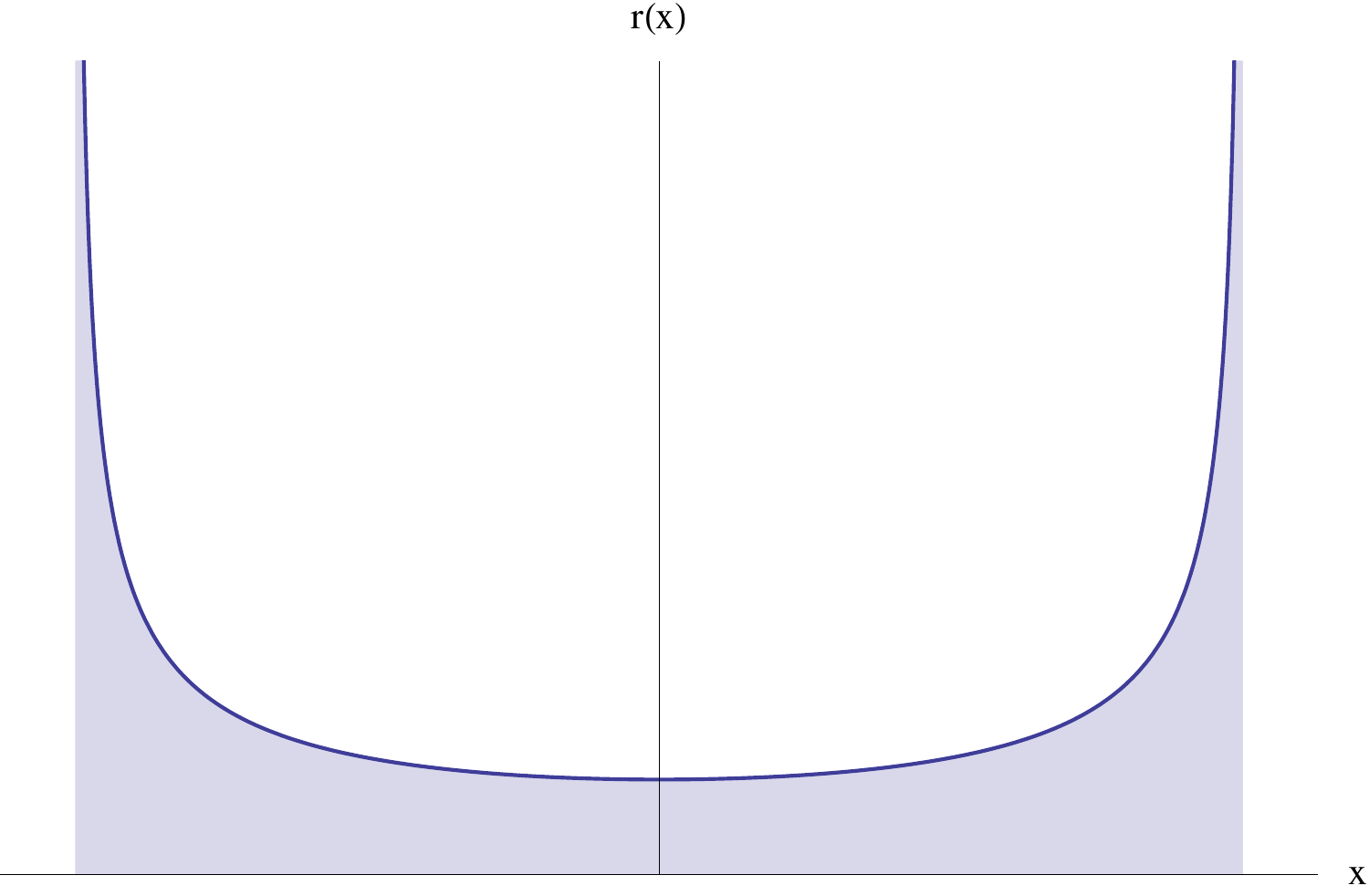}
   \label{fig2:subfig1}
 }
 \subfigure[]{
   \includegraphics[height=4.2cm, trim=0cm 0.7cm 0cm 0cm] {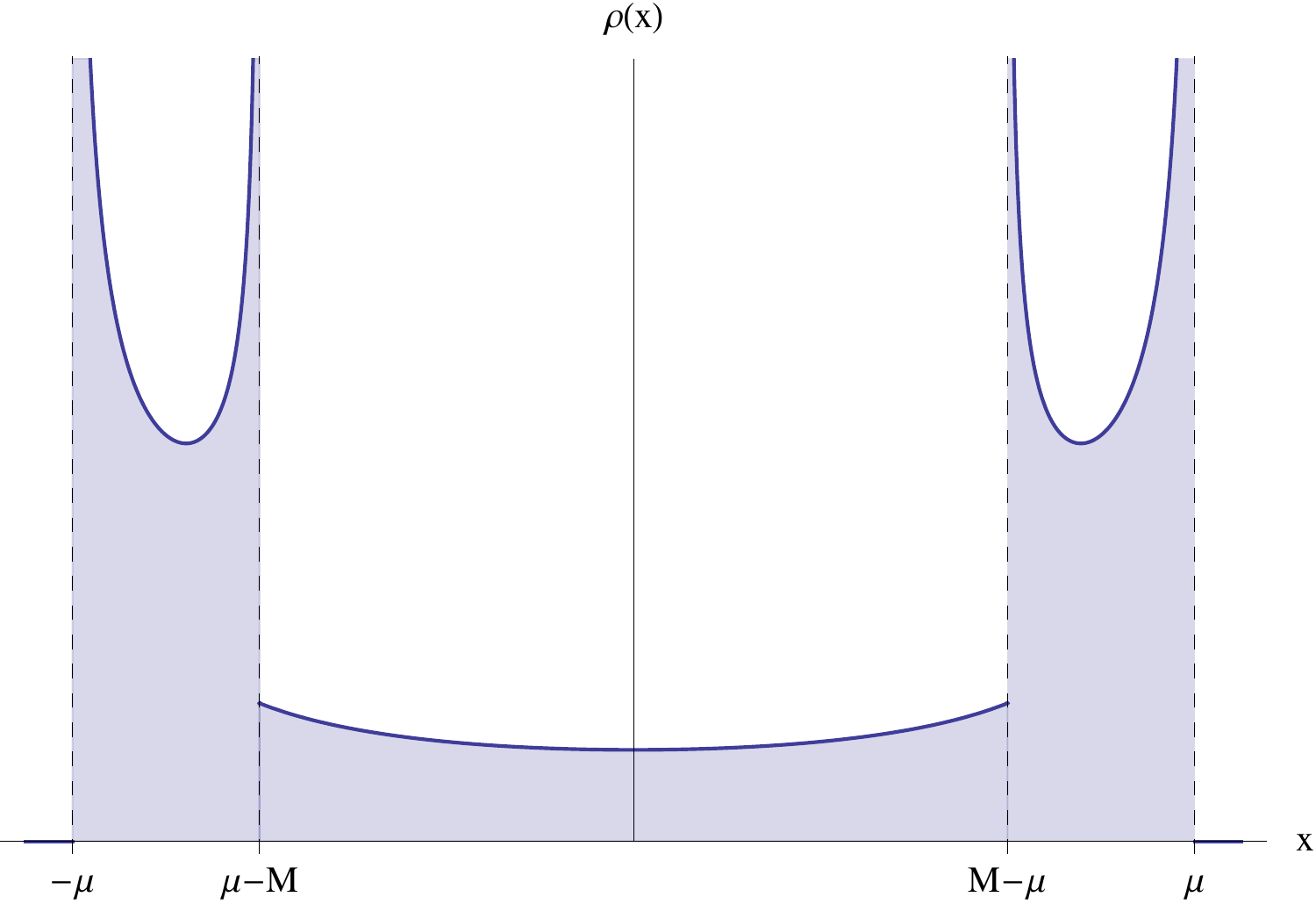}
   \label{fig2:subfig2}
 }
\caption{\label{fig1}\small The non-local map (\ref{rho-r}) induces additional singularities inside the interval $(-\mu ,\mu )$: (a) the auxiliary function $r(x)$; (b) and its image $\rho (x)$.}
\end{center}
\end{figure}

The coefficients in (\ref{rho-r}) are adjusted in way that guarantees cancellation of the resonance terms (those terms which have poles at $y=x\pm M$) in the integral equation (\ref{intnormal1}). Indeed upon substituting the ansatz (\ref{rho-r}) into the saddle-point equation (\ref{intnormal1}), many different terms cancel.
Collecting what is left,  we get an integral equation of the Hilbert type:
\begin{eqnarray}
 \strokedint_{-\mu }^\mu  dy\,r(y)\,\frac{2}{x-y}
 &=&\int_{-M+\mu }^{M-\mu } dy\,r (y)\left(\frac{1}{x-y-M}+\frac{1}{x-y+M}\right)
\nonumber \\
 &&
 +\frac{2}{3}\int_{M-\mu }^{\mu }
 dy\,r (y)\left(\frac{2}{x-y-M}+\frac{1}{x-y+2M}\right)
 \nonumber \\
 &&
+\frac{2}{3}\int_{-\mu }^{\-M+\mu }
 dy\,r (y)\left(\frac{2}{x-y+M}+\frac{1}{x-y-2M}\right).
\end{eqnarray}
All three integrals on the right hand side are proper and consequently define a smooth function on  $(-\mu ,\mu )$.  Basic theory of singular integral equations \cite{Gakhov} guarantees the existence of solution with the boundary conditions
\begin{equation}\label{rasympt}
 r(x)\rightarrow \frac{1}{\sqrt{\mu \mp x}}\qquad \left(x\rightarrow \pm \mu \right).
\end{equation}
Notice that here we fix the constant in front of the square root, which removes the  rescaling ambiguity. The solution is thus unique and has a shape illustrated in fig~\ref{fig2:subfig1}.

From the ansatz (\ref{rho-r}) we then find that the density behaves as
\begin{equation}
 \rho (x)\rightarrow \frac{4}{3C\sqrt{\mu \mp x}}\qquad \left(x\rightarrow \pm \mu \right)
\end{equation}
at the boundaries of the interval. When $x$ approaches one of the resonance points from the outside, the density diverges as
\begin{equation}
 \rho (x)\rightarrow \frac{2}{3C\sqrt{ \pm x -M+\mu }}\qquad \left(x\rightarrow \pm M\mp\mu \pm 0 \right).
\end{equation}
On the inside, the density approaches a finite value $r(M-\mu )/C$. This structure is illustrated in fig.~\ref{fig2:subfig2}.

Finally, the constant $C$ is determined by the normalization condition. In terms of the auxiliary function $r(x)$, the normalization condition reads:
\begin{equation}\label{norC}
 \left(\int_{-\mu }^{\mu }+\int_{-\mu }^{-M+\mu }+\int_{M-\mu }^{\mu }\right)
 dx\,r(x)=C.
\end{equation}

\section{General structure}

More resonances appear with growing $\mu $, leading to secondary phase transitions each time $2\mu$ passes an integer multiple of $M$. In the $n$-th phase the eigenvalue density has $2n$ spikes inside the interval $(-\mu ,\mu )$. To characterize different phases, it is convenient to introduce two variables \cite{Anderson:2014hxa}:
\begin{equation}
 n=\left[\frac{2\mu }{M}\right],\qquad \Delta =\left\{\frac{2\mu }{M}\right\},\qquad 2\mu =(n+\Delta)M .
\end{equation}
where $[]$ and $\{\}$ denote integer and fractional parts, respectively.

\begin{figure}[t]
\begin{center}
 \centerline{\includegraphics[width=12cm]{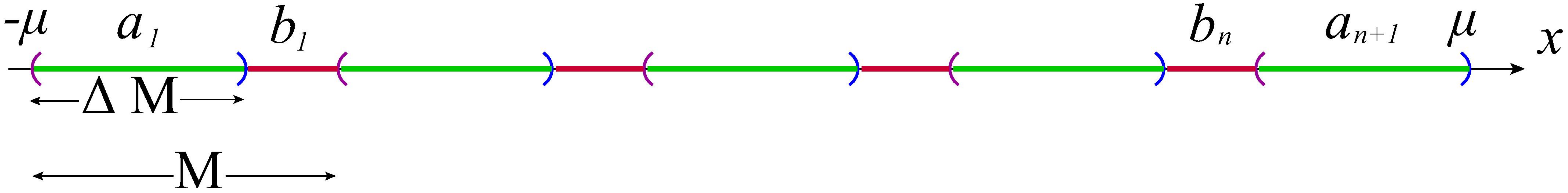}}
\caption{\label{a-bfig}\small The  resonance points divide the eigenvalue interval into subintervals $a_k$ and $b_k$, on which the density is non-singular.}
\end{center}
\end{figure}

The spikes are located at the resonance points $-\mu +kM$ and $\mu -kM=-\mu +(n-k+\Delta)M $, which divide the interval $(-\mu,\mu )$ into $2n+1$ subintervals,   $a_k$ and $b_k$ \cite{Anderson:2014hxa}, as shown in fig.~\ref{a-bfig}:
\begin{eqnarray}
  &a_k=\left(-\mu +(k-1)M,-\mu +(k-1+\Delta)M \right), & {\rm }k=1,\ldots ,n+1
\\
    &b_k=(-\mu +(k-1+\Delta)M ,-\mu +kM ),{\rm ~~~~~~~} &{\rm }k=1,\ldots ,n.
\end{eqnarray}
The key assumption we make is that the density can be obtained from a unique analytic function $r(x)$ by applying an appropriate shift operator. We thus take the following ansatz that generalizes (\ref{rho-r}) to the case of arbitrary number of subintervals:
\begin{equation}\label{rho-many-r}
 C\rho (x)=
\begin{cases}
 \sum\limits_{l=-k+1}^{n+1-k}A_{k+l,l}\,r\left(x+lM\right), & {\rm }x\in a_k
\\
 \sum\limits_{l=-k+1}^{n-k}B_{k+l,l}\,r\left(x+lM\right), & {\rm }x\in b_k.
\end{cases}
\end{equation}
The numerical coefficients  $A_{mp}$, $B_{mp}$ are to be determined later.

After substituting this ansatz into the saddle-point equation (\ref{intnormal1}), we get a linear combination of integrals of the form
\begin{equation}
 I_{mp}=\strokedint_{a_k}\frac{dy\,r\left(y\right)}{x-y+pM}\,,\qquad 
  J_{mp}=\strokedint_{b_k}\frac{dy\,r\left(y\right)}{x-y+pM}\,.
\end{equation}
More precisely,
\begin{eqnarray}
 &&\strokedint_{-\mu }^\mu
 dy\,\rho (y)  \left( \frac{2}{x-y} - \frac{1}{x-y+M}-\frac{1}{x-y-M} \right) 
 \nonumber \\
&&
 =
\sum_{m=1}^{n+1}\sum_{l=m-n-1}^{m-1}A_{ml}\left(2I_{ml}-I_{m,l+1}-I_{m,l+1}\right)
 \nonumber \\
&&
+\sum_{m=1}^{n}\sum_{l=m-n}^{m-1}B_{ml}\left(2J_{ml}-J_{m,l+1}-J_{m,l+1}\right).
\nonumber 
\end{eqnarray}

This  results in an integral equation for $r(x)$. The goal is to cancel all singular terms in this equation (except for those with the Hilbert kernel) by adjusting the coefficients $A_{mp}$ and $B_{mp}$. The equation will then be of the Hilbert type and the existence of solution will be guaranteed by general theorems.
The dangerous integrals to cancel are $I_{mp}$ with $m-n-1\leq p\leq m-1$, $p\neq 0$, and $J_{mp}$ with $m-n\leq p\leq m-1$, $p\neq 0$. Imposing the condition that the coefficients in front of these integrals vanish, and normalizing the coefficient of the Hilbert term to 2, we get a system of linear equations on $A_{mp}$, $B_{mp}$:
\begin{eqnarray}
 &2A_{mp}-A_{m,p+1}-A_{m,p-1}=2\delta _{p0},& p=m-n-1,\ldots ,m-1
\\
&2B_{mp}-B_{m,p+1}-B_{m,p-1}=2\delta _{p0},& p=m-n,\ldots ,m-1
\end{eqnarray}
The solution with the correct boundary conditions is
\begin{equation}\label{Amp}
 A_{mp}=
\begin{cases}
 \frac{2\left(n+2-m\right)\left(m-p\right)}{n+2}\,, & {\rm }p\geq 0
\\
  \frac{2m\left(n+2-m+p\right)}{n+2}\,, & {\rm }p\leq 0
\end{cases}
\end{equation}
and
\begin{equation}\label{Bmp}
B_{mp}=
\begin{cases}
 \frac{2\left(n+1-m\right)\left(m-p\right)}{n+1}\,, & {\rm }p\geq 0
\\
  \frac{2m\left(n+1-m+p\right)}{n+1}\,, & {\rm }p\leq 0.
  \end{cases}
\end{equation}
An easy check is to  see that (\ref{rho-r}) is reproduced when  $n=1$.
 
The resulting integral equation for $r(x)$ takes the form:
\begin{eqnarray}\label{s-int-eq-for-r}
 \strokedint_{-\mu }^\mu  dy\,\,\frac{r(y)}{x-y}
 &=&\frac{1}{n+2}\sum_{m=1}^{n+1}
 \int_{a_m }
 dy\,r (y)\left[\frac{m}{x-y-\left(n+2-m\right)M}+\frac{n+2-m}{x-y+mM}\right]
 \nonumber \\
 &&
+\frac{1}{n+1}\sum_{m=1}^{n}\int_{b_m }
 dy\,r (y)\left[\frac{m}{x-y-\left(n+1-m\right)M}+\frac{n+1-m}{x-y+mM}\right].
\end{eqnarray}
The right-hand side is indeed analytic on the interval $\left(-\mu ,\mu \right)$. Hence the solution to this equation with the asymptotics (\ref{rasympt}) exists and is unique.
The analog of equation (\ref{norC}) now reads
\begin{equation}\label{Cnor}
\left[\sum_{m=1}^{n+1}m\left(n+2-m\right)\int_{a_m}^{}+\sum_{m=1}^{n}m\left(n+1-m\right)\int_{b_m}^{}\right]dx\,r(x)=C.
\end{equation}
This condition determines the normalization constant $C$ once the function $r(x)$ is known.

Generic average, for instance the one that determines the coupling constant as a function of $\mu $ via (\ref{formintegrated}), can be computed as follows. For a function $\mathcal{F}(x)$, define a projection on the $m$-th interval: 
\begin{eqnarray}\label{projection}
 \hat{\mathcal{F}}_{ma}(x)&=&\frac{2m}{n+a}\sum_{l=0}^{n+a-1-m}(n+a-m-l)\mathcal{F}(x+lM)
\nonumber \\
&&+\frac{2(n+a-m)}{n+a}\sum_{l=1}^{m-1}(m-l)\mathcal{F}(x-lM),
\end{eqnarray}
where  $a=1,2$ distinguishes  between the $a$ and $b$ intervals. In these notations,
\begin{equation}\label{rho-many-r}
 C\rho (x)=
\begin{cases}
 \hat{r}_{m2}(x), & {\rm }x\in a_m
\\
  \hat{r}_{m1}(x), & {\rm }x\in b_m.
\end{cases}
\end{equation}
The average can thus be calculated by applying the projection to $r(x)$. Alternatively, the projection can be applied to the function itself:
\begin{equation}\label{unnornalized}
 \left\llangle \mathcal{F}(x)\right\rrangle=\sum_{m=1}^{n+1}\int_{a_m}^{}dx\,r(x)\hat{\mathcal{F}}_{m2}(x)+\sum_{m=1}^{n}\int_{b_m}^{}dx\,r(x)\hat{\mathcal{F}}_{m1}(x).
\end{equation}
This defines an unnormalized average of $\mathcal{F}(x)$. 
The normalized expectation value is obtained by dividing with $\llangle 1\rrangle$:
\begin{equation}
 \left\langle \mathcal{F}(x)\right\rangle=\frac{\left\llangle\mathcal{F}(x) \right\rrangle}{\left\llangle 1\right\rrangle}\,.
\end{equation}

For instance,
\begin{equation}\label{onehat}
 \hat{1}_{ma}=m(n+a-m), 
\end{equation}
so the condition (\ref{Cnor}) is indeed equivalent to
\begin{equation}
\left\llangle 1\right\rrangle = C,
\end{equation}
guaranteeing that $\langle 1\rangle=1$. For $\left\langle x^2\right\rangle$ that determines the  vacuum susceptibility (\ref{susceptibility}), we get:
\begin{eqnarray}\label{xwithahat}
 \widehat{x^2}_{ma}&=&m\left(n+a-m\right)\left\{
 \left[x+\frac{M}{3}\left(n+a-2m\right)\right]^2
\right.\nonumber \\ &&\left.
+\frac{M^2}{18}\left[
 \left(n+a\right)^2-m\left(n+a\right)+m^2-3
 \right]
 \right\}.
\end{eqnarray}

It is convenient to introduce a dimensionless variable
\begin{equation}
 \xi =\frac{x+\mu }{M}-m+1\,,
\end{equation}
that takes values from $0$ to $\Delta $ on $a_m$ and from $\Delta $ to $1$ on $b_m$, and regard (\ref{projection}) as a function of $\xi $:
\begin{eqnarray}\label{xi-proj}
 \hat{\mathcal{F}}_{ma}(\xi )&=&\frac{2}{n+a}\sum_{l=1}^{n+a-1}\left[
 \theta (l-m)m(n+a-l)+\theta (m-l)l(n+a-m)
 \right]
\nonumber \\
&&\times 
 \mathcal{F}\left(-\mu +(l-1+\xi )M\right).
\end{eqnarray}
The average (\ref{unnornalized}) can be compactly written as
\begin{equation}\label{compact-average}
 \left\llangle\mathcal{F}(x) \right\rrangle=M\int_{0}^{1}d\xi \,
 \sum_{m=1}^{n+\theta (\Delta -\xi )}r\left(-\mu +(m-1+\xi )M\right)
 \hat{\mathcal{F}}_{m,1+\theta (\Delta -\xi )}(\xi ).
\end{equation}

For example, defining
\begin{equation}
 F(x)=\ln\frac{\left|M^2-(\mu +x)^2\right|}{(\mu +x)^2}\,,
\end{equation}
we find:
\begin{equation}\label{Fhat}
 \hat{F}_{ma}(\xi )=2\ln\frac{1-\xi }{\xi +m-1}+\frac{2m}{n+a}\,\ln\frac{\xi +n+a-1}{1-\xi }\,.
\end{equation}
The condition (\ref{formintegrated}) now becomes
\begin{equation}\label{coupling-}
 \frac{8\pi ^2}{\lambda }=\frac{\left\llangle F(x)\right\rrangle}{\left\llangle 1 \right\rrangle}\,,
\end{equation}
and can be easily handled with the help of (\ref{onehat}) and (\ref{Fhat}).

\begin{figure}[t]
\begin{center}
 \centerline{\includegraphics[width=8cm]{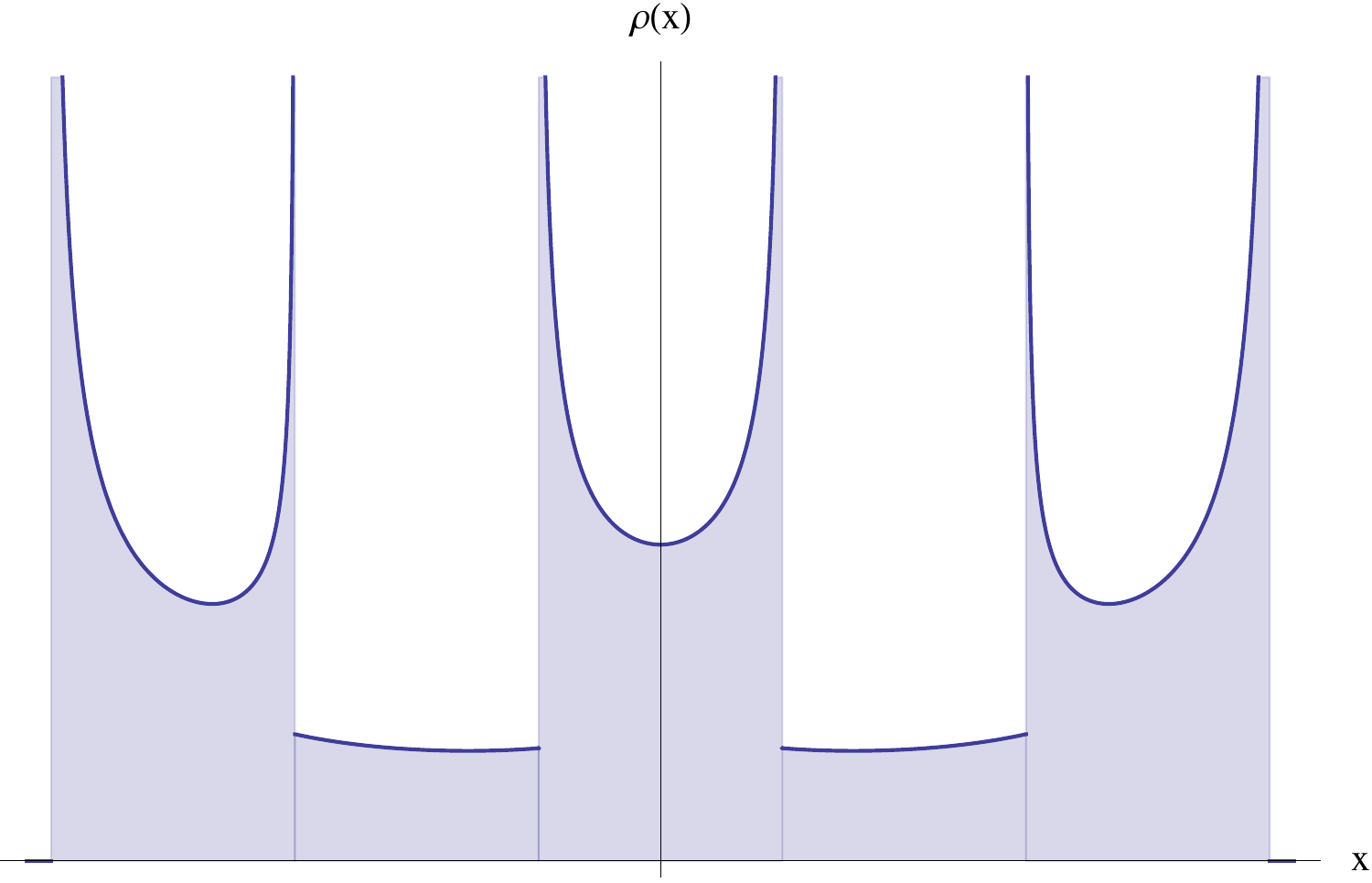}}
\caption{\label{fig-n2}\small The eigenvalue density for $n=2$.}
\end{center}
\end{figure}

To summarize, the density, the coupling constant and any matrix-model average are all determined through one universal function $r(x)$, which is a solution of the singular integral equation (\ref{s-int-eq-for-r}) with the boundary conditions (\ref{rasympt}). While we do not know how to solve this equation analytically, a
numerical solution can be easily obtained by iterations.
The salient features of the density can be actually inferred from the map (\ref{rho-many-r}) alone, without knowing the detailed form of the function $r(x)$. This concerns, for instance, the structure of singularities. As can be seen from the definition of the map (\ref{rho-many-r}), the density is non-singular on all the $b$-intervals, and has singularities at  both ends of  the $a$-intervals. The relative strength of these singularities is completely determined by the $r(x)\rightarrow \rho (x)$ map and the boundary condition on $r(x)$:
\begin{eqnarray}
 \rho (x)&\simeq&\frac{2\left(n+1-l\right)}{C\left(n+2\right)\sqrt{x+\mu -lM }}\,,\qquad x\rightarrow -\mu +lM +0
\\
 \rho (x)&\simeq&\frac{2l}{C\left(n+2\right)\sqrt{lM+\Delta -\mu -x}}\,,\qquad x\rightarrow -\mu +lM+\Delta -0. 
 \end{eqnarray}
The structure of singularities is illustrated in fig.~\ref{fig-n2}.

\section{Critical behavior}

\begin{figure}[t]
\begin{center}
 \subfigure[]{
   \includegraphics[height=4.15cm, trim=0cm 0cm 0cm 0cm] {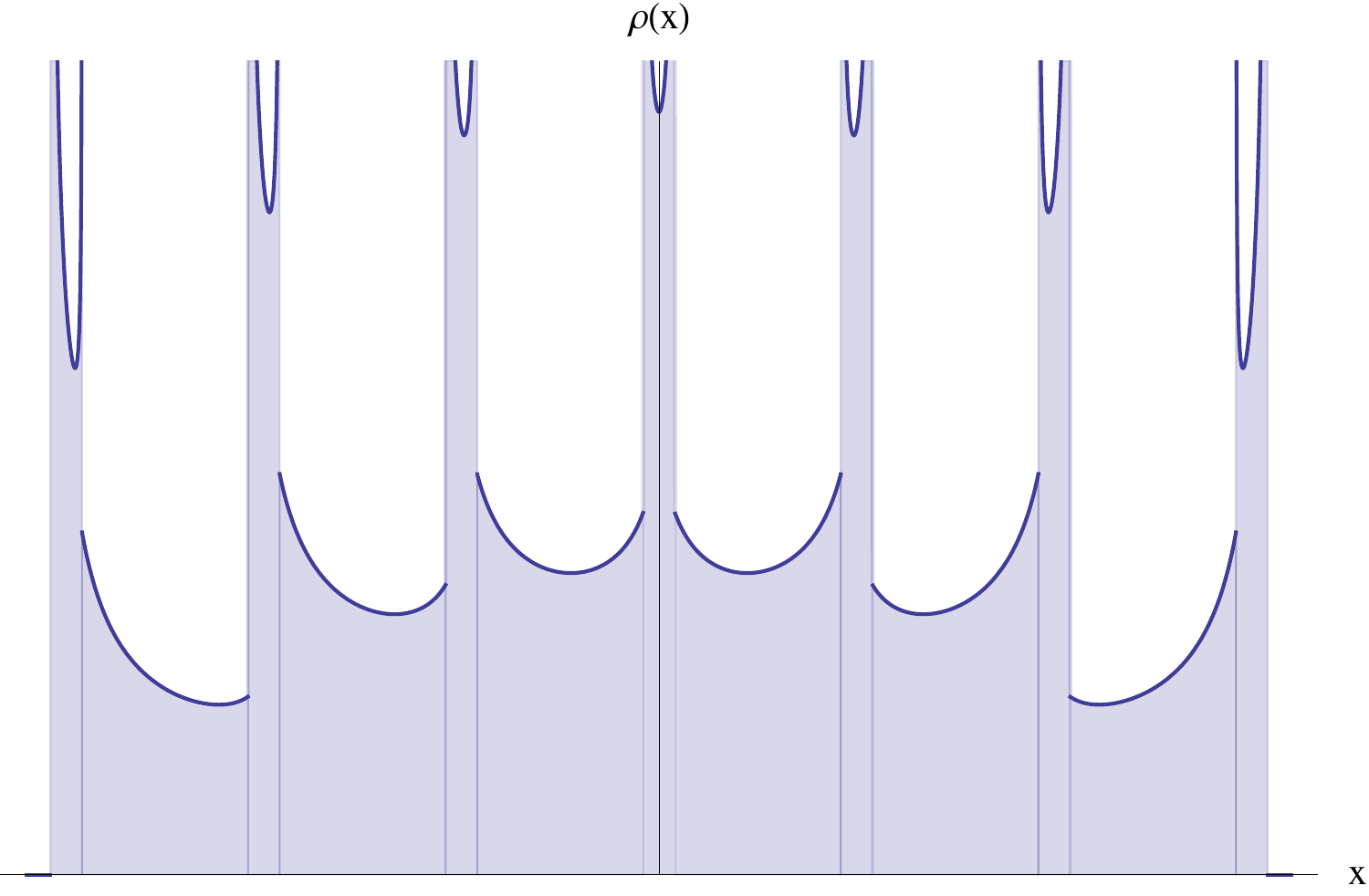}
   \label{fig-last:subfig1}
 }
 \subfigure[]{
   \includegraphics[height=4.15cm, trim=0cm 0cm 0cm 0cm] {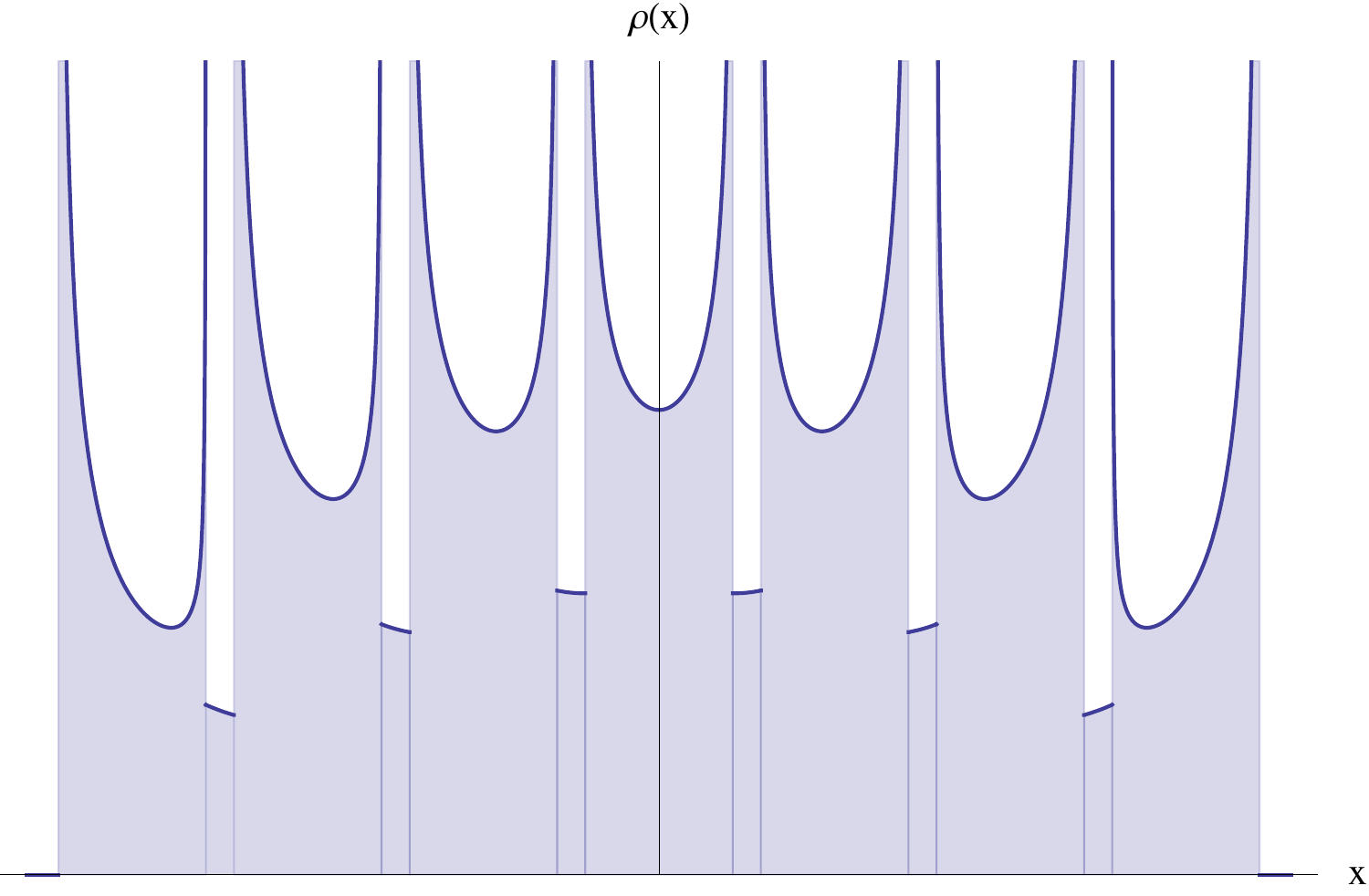}
   \label{fig-last:subfig2}
 }
\caption{\label{fig-last}\small The critical behavior in $\mathcal{N}=2^*$ SYM: (a) when the critical point is approached from above, the $a$-intervals shrink to zero size ($\Delta \rightarrow 0$); (b) when the critical point is approached from below, the $b$-intervals shrink ($\Delta \rightarrow 1$).}
\end{center}
\end{figure}

When $\Delta $ approaches one from below, $\Delta \rightarrow 1^-$, $n$ jumps by one unit and the system undergoes a transition to a new phase in which $\Delta $ starts growing again from zero. Similarly, $\Delta \rightarrow 0^+$ corresponds to approaching the critical point from above.  In each case, either $b$ or $a$-intervals  shrink to a point, as in fig.~\ref{fig-last}. Exactly at the critical point a distinction between $a$ and $b$ interval disappears. Moreover, the boundary conditions (\ref{rasympt}) cannot be imposed any more and the endpoint behavior of the density is governed by a non-trivial critical exponent.

To study the critical behavior, it is convenient to invert the Hilbert kernel in the integral equation (\ref{s-int-eq-for-r}). First we notice that the equation can be compactly written as
\begin{equation}\label{inteq-compact}
 \strokedint_{-\mu }^\mu  dy\,\,\frac{r(y)}{x-y}
 =-Mx\int_{0}^{1}d\xi \,\,\frac{\hat{r}_{1,1+\theta (\xi-1+\Delta  )}(1-\xi )}{(\mu +M\xi )^2-x^2}\,,
\end{equation}
where $\hat{r}_{ma}$ was introduced  in (\ref{xi-proj}). Solving for $r(x)$, we get:
\begin{equation}\label{r-intrep}
 r(x)=\frac{M^{\frac{3}{2}}}{\pi \sqrt{\mu ^2-x^2}}
 \int_{0}^{1}d\xi \,\hat{r}_{1,1+\theta (\xi-1+\Delta  )}(1-\xi )\,
 \frac{(\mu +M\xi )\sqrt{\xi (2\mu +M\xi )}}{(\mu +M\xi )^2-x^2}\,.
\end{equation}
This is just another form of the integral equation -- the right-hand side still depends on the unknown function through the non-local map (\ref{xi-proj}). But this form is far more convenient than (\ref{inteq-compact}) for iterative numerical solution, as well as for perturbative analysis at strong coupling, to be discussed later. 

\begin{figure}[t]
\begin{center}
 \centerline{\includegraphics[width=10cm]{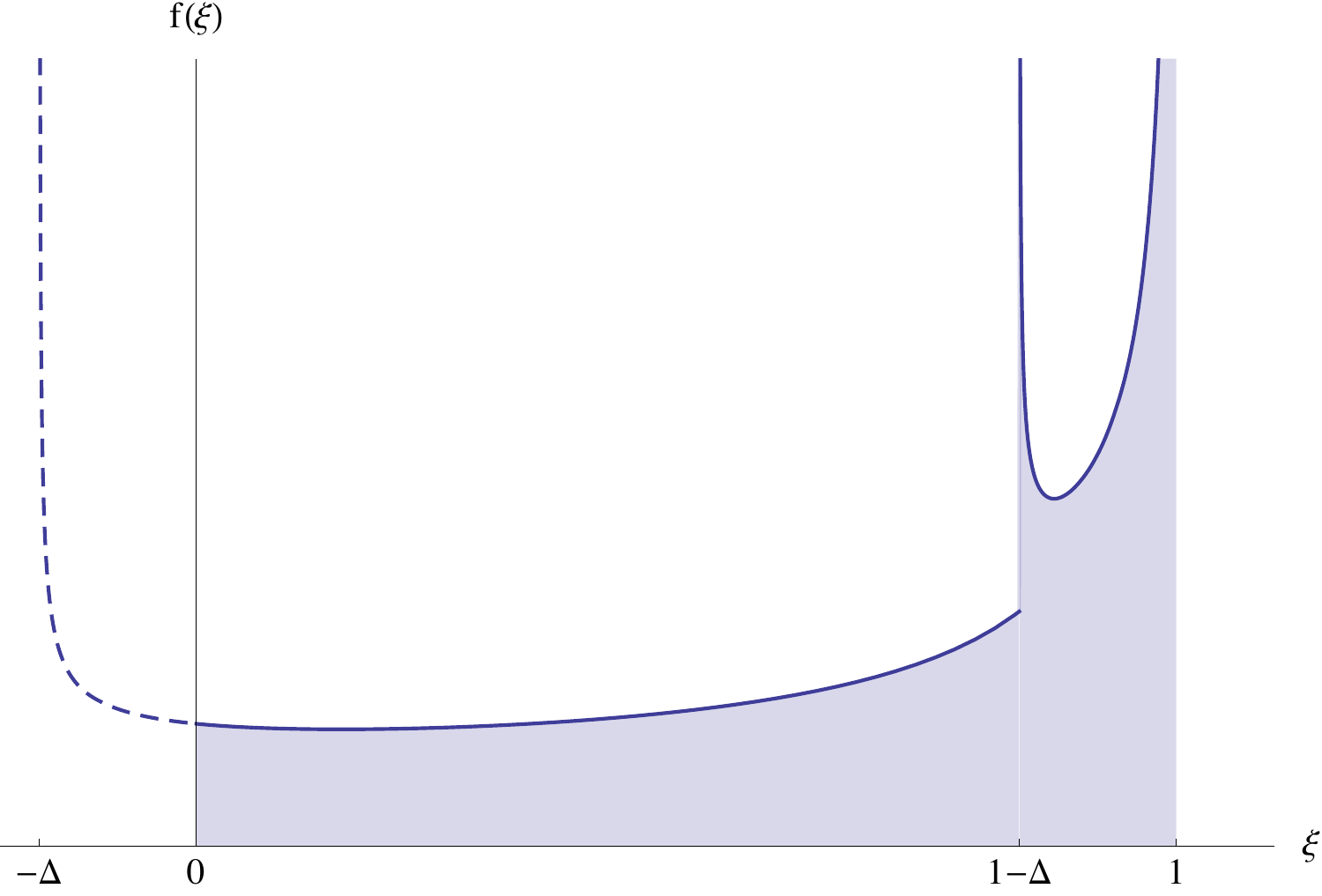}}
\caption{\label{under}\small The function $f(\xi )\equiv \hat{r}_{1,1+\theta (\xi-1+\Delta  )}(1-\xi )$. The choice of parameters is the same as in fig.~\ref{fig-last:subfig1}.}
\end{center}
\end{figure}

The function $\hat{r}_{1,1+\theta (\xi-1+\Delta  )}(1-\xi )$ that enters the integral representation (\ref{r-intrep}) has a shape illustrated in fig.~\ref{under}. It has integrable singularities at $\xi =1$ and $\xi =1-\Delta $, which are images of the endpoint singularity of $r(x)$, but is regular at zero. This is crucial for consistency of the boundary conditions. Indeed the integral in (\ref{r-intrep}) converges well at zero and as a consequence defines an analytic function in the vicinity of the cut $(-\mu ,\mu )$, which guarantees the correct boundary conditions (\ref{rasympt}). 

However, the function in fig.~\ref{under} has a shadow singularity at $\xi =-\Delta $, outside of the integration domain in (\ref{r-intrep}). When the critical point is approached from above, $\Delta \rightarrow 0$, and the singularity hits the integration region. Exactly at the critical point, the integrand becomes singular at zero invalidating the boundary conditions (\ref{rasympt}).  The main contribution to the integral then comes from small $\xi $ and, if $x$ is close to $\mu $ we get:
\begin{equation}\label{small-eta-eq}
 r(\mu -M\eta )\simeq \frac{1}{2\pi \sqrt{\eta }}\int_{0}^{}d\xi \,
 \hat{r}_{11}(1-\xi )\,\frac{\sqrt{\xi }}{\xi +\eta }\,.
\end{equation}
Since the boundary conditions (\ref{rasympt}) cannot be imposed any more, the edge-point behavior of the density changes.
Assuming that the modified asymptotics is power-like:
\begin{equation}
 r(\mu -M\eta )\simeq \frac{A}{\eta ^\alpha }\,
\end{equation}
we find that
\begin{equation}
  \hat{r}_{11}(1-\xi )\simeq \frac{2}{n+1}\,r(\mu -M\xi )\simeq \frac{2A}{(n+1)\xi ^\alpha }\,,
\end{equation}
because the sum in (\ref{xi-proj}) receives the main contribution from the last term. Substituting this back into (\ref{small-eta-eq}), we get a consistency condition on the critical exponent $\alpha $:
\begin{equation}
 \frac{1}{\eta ^\alpha }\simeq \frac{1}{\pi (n+1)\sqrt{\eta }}\int_{0}^{}d\xi \,\,
 \frac{\xi ^{\frac{1}{2}-\alpha }}{\xi +\eta }\simeq 
 \frac{1}{(n+1)\sin\left(\pi \alpha -\frac{\pi }{2}\right) }\,\,\frac{1}{\eta ^\alpha }\,,
\end{equation}
The critical exponent at the $n$-th phase transition consequently is equal to
\begin{equation}
 \alpha =\frac{1}{2}+\frac{1}{\pi }\,\arcsin\frac{1}{n+1}\,.
\end{equation}
For $n=1$, we get $\alpha =2/3$, reproducing the result of \cite{Russo:2013kea} for the critical exponent at the weak-strong coupling phase transition. The transitions between different strong-coupling phases ($n>1$) are milder, in the sense that at larger  $n$ the critical index becomes closer and closer to $1/2$, the unperturbed endpoint exponent  away from criticality.

\section{Strong coupling}

\begin{figure}[t]
\begin{center}
 \centerline{\includegraphics[width=12cm]{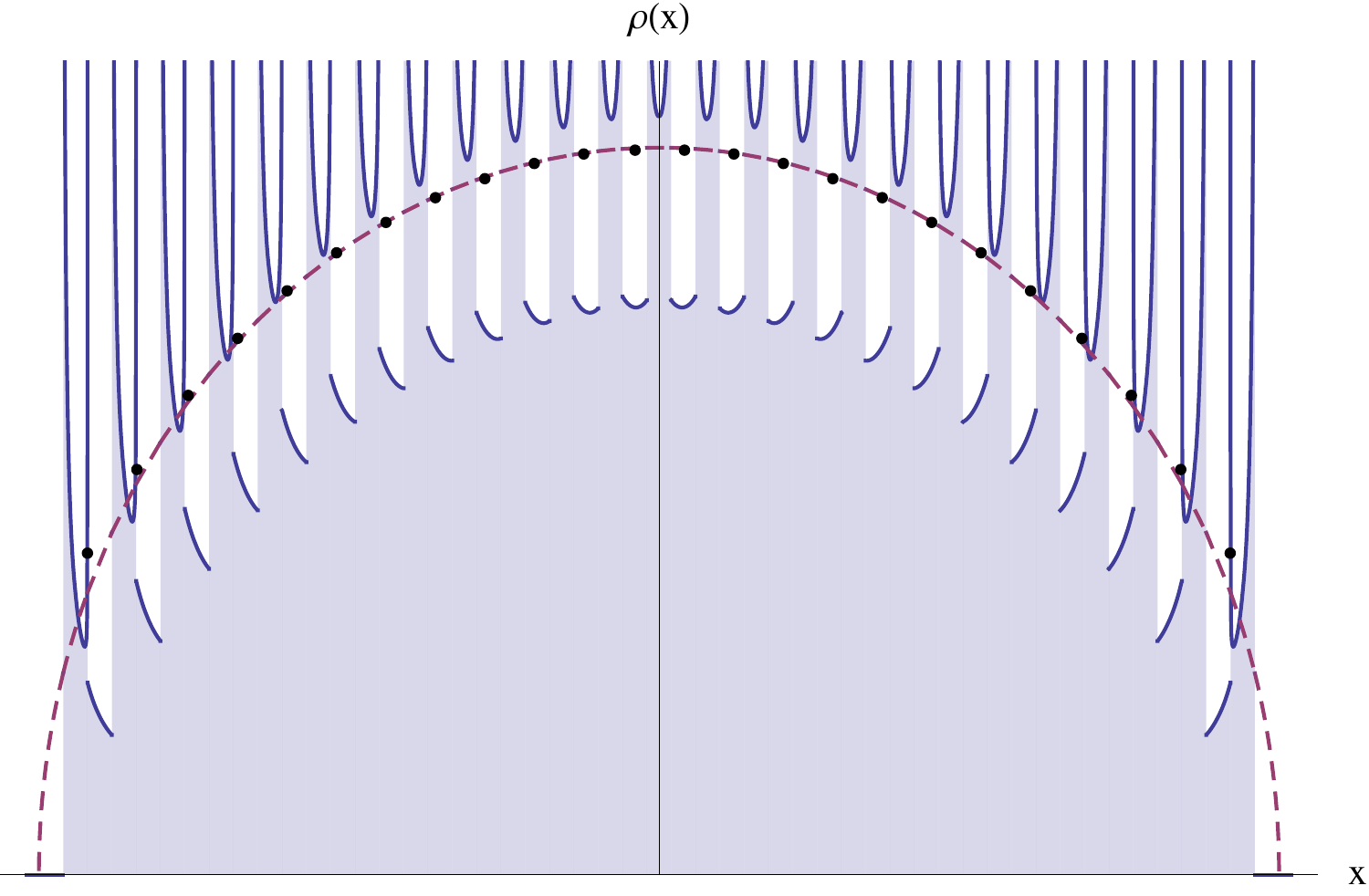}}
\caption{\label{wigner}\small The eigenvalue density at strong coupling, computed numerically. On small scales the density has a complicated spiky structure with a large number of singularities. However, the averaged density (black dots) is well approximated by the Wigner semicircle, shown in the dashed line.}
\end{center}
\end{figure}

The strong-coupling regime of the planar $\mathcal{N}=2^*$ SYM is of particular interest, because of possible connections to the holographic duality. The holographic dual of $\mathcal{N}=2^*$ theory, known as the Pilch-Warner background \cite{Pilch:2000ue}, is an asymptotically $AdS_5\times S^5$ solution of type IIB supergravity with a domain wall in the bulk. The density of Higgs eigenvalues  (\ref{eigenvalue-density}) can be calculated holographically from the D-brane probe analysis of this background, with the result that \cite{Buchel:2000cn}
\begin{equation}\label{rho-Wigner}
 \rho _\infty (x)=\frac{2}{\pi \mu ^2}\,\sqrt{\mu ^2-x^2}\,,
\end{equation}
where the width of the eigenvalue distribution is related to the 't~Hooft coupling and the mass scale of the dual SYM theory as
\begin{equation}\label{string-lambda}
 \mu =\frac{\sqrt{\lambda }M}{2\pi }\,.
\end{equation}
This prediction is supposed to be accurate at large $\lambda $, up to possible $1/\sqrt{\lambda }$ corrections, where $\sqrt{\lambda }/2\pi$ is identified with the string tension measured in the curvature units.

On the matrix model side, the width of the eigenvalue distribution  becomes parametrically larger than the mass scale  at strong coupling: $\mu \gg M$. In that case one can use a simple approximate scheme to solve the saddle-point equations  \cite{Buchel:2013id}. The resulting solution reproduces the Wigner density (\ref{rho-Wigner}) and its correct width\footnote{An alternative derivation of the semicircle law that does not rely on $S^4$ localization is given in \cite{Billo:2014bja}.} thus confirming the predictions of holography. Moreover, perimeter law for the Wilson loop vev (\ref{Perimeter}) can be matched to the area law in the dual geometry \cite{Buchel:2013id}.

These simple results almost contradict the preceding discussion, from which we may expect that the density becomes increasingly more complicated at bigger and bigger $\mu $.
Indeed, when  $\mu \gg M$, the number of intervals $2n+1$ gets very large, and the density acquires an irregular, spiky structure, with an increasing number of singularities. However, upon averaging over sufficiently large interval the small-scale irregularities are smoothened out, and one indeed finds a nearly perfect agreement with the simple Wigner distribution, as illustrated in fig.~\ref{wigner}.  We will now  develop a systematic strong-coupling expansion for the eigenvalue density, which as we shall see reproduces the Wigner distribution  at the leading order and at the next-to-leading order agrees with the results obtained in \cite{Chen:2014vka}. The infinite-volume ($R\rightarrow \infty $) and strong-coupling ($\lambda \rightarrow \infty $) limits were taken in \cite{Chen:2014vka} in the opposite order. The agreement that we find means that there are no order-of-limits ambiguities, in spite that $(R,\lambda )=(\infty ,\infty )$ is an accumulation point of an infinite number of phase transitions.

At strong coupling when $x\sim \mu \gg M$, we can compute the master function $r(x)$ by expanding (\ref{r-intrep}) in power series in $M $:
\begin{equation}\label{expansioninKs}
 r(x)=\frac{M^2\mathbbm{K}_0}{\pi (\mu ^2-x^2)^{\frac{3}{2}}}
 -\frac{M^3\mathbbm{K}_1(3\mu ^2+5x^2)}{4\pi \mu (\mu ^2-x^2)^{\frac{5}{2}}}+\ldots ,
\end{equation}
where the coefficients $\mathbbm{K}_n$ are defined as
\begin{equation}\label{KK's}
 \mathbbm{K}_n=\sqrt{\frac{2\mu ^3}{M}}\int_{0}^{1}d\xi \,\hat{r}_{1,1+\theta (\xi-1+\Delta  )}(1-\xi )
 \xi ^{n+\frac{1}{2}}.
\end{equation}

Since the sizes of $a_m$ and $b_m$ intervals are much smaller than the total width of the eigenvalue distribution, $m$ and $n$ in the projection operator (\ref{xi-proj}) can  be replaced by $(x+\mu )/M$ and $2\mu /M$. Summation over $l$ becomes integration and we get:
\begin{equation}
 \hat{\mathcal{F}}(x)\simeq \frac{1}{\mu M^2}\int_{-\mu }^{\mu }dy\,
 \left[
 \theta (y-x)\left(\mu +x\right)\left(\mu -y\right)
 +\theta (x-y)\left(\mu -x\right)\left(\mu +y\right)
 \right]\mathcal{F}(y).
\end{equation}
We can now compute the density from (\ref{rho-many-r}), by applying this operator to (\ref{expansioninKs}). Keeping only the first term, we get:
\begin{equation}\label{leadingrho}
 \rho (x)=\frac{1}{C}\,\hat{r}(x)\simeq \frac{2\mathbbm{K}_0}{\pi \mu ^2C}\,\sqrt{\mu ^2-x^2}.
\end{equation}
For the normalization constant we find:
\begin{equation}\label{constant_C}
 C=\left\llangle 1\right\rrangle\simeq \frac{1}{M^2}\int_{-\mu }^{\mu }dx\,\left(\mu ^2-x^2\right)r(x)\simeq \mathbbm{K}_0,
\end{equation}
where we used (\ref{unnornalized}) and (\ref{onehat}) and again replaced summation with integration. Together with (\ref{leadingrho}), this
reproduces the leading-order Wigner distribution (\ref{rho-Wigner}).

To compute the coupling constant from (\ref{Fhat}), (\ref{coupling-}), we first need to average over $\xi $:
\begin{equation}
 \left\llangle F(x)\right\rrangle\simeq \frac{1}{\mu }\int_{-\mu }^{\mu }dx\,
 r(x)
 \int_{0}^{1}d\xi \,\left[\left(\mu -x\right)\ln\frac{M(1-\xi )}{2\mu }+2\mu \ln\frac{2\mu }{x+\mu }\right]
 \simeq \frac{2M^2\mathbbm{K}_0}{\mu ^2}\,,
\end{equation}
and we then get:
\begin{equation}
 \frac{8\pi ^2}{\lambda }=\frac{\left\llangle F(x)\right\rrangle}{\left\llangle 1\right\rrangle}=\frac{2M^2}{\mu ^2}\,,
\end{equation}
in complete agreement with (\ref{string-lambda}).

It is quite straightforward to include the next term in the expansion (\ref{expansioninKs}). We then get:
\begin{eqnarray}\label{various-averages}
 \left\llangle 1\right\rrangle&=&\mathbbm{K}_0+\frac{5\mathbbm{K}_1M}{4\mu }+\ldots 
\nonumber \\
\left\llangle F(x) \right\rrangle&=&\frac{2M^2}{\mu ^2}\left(\mathbbm{K}_0-\frac{\mathbbm{K}_1M}{12\mu }+\ldots \right),
\end{eqnarray}
thus
\begin{equation}
 \frac{4\pi ^2}{\lambda }=\frac{M^2}{\mu ^2}\left(1-\frac{4\mathbbm{K}_1M}{3\mathbbm{K}_0\mu }+\ldots \right),
\end{equation}
and
\begin{equation}\label{effstringT}
 \mu =\frac{\sqrt{\lambda }M}{2\pi }\left(1-\frac{4\pi \mathbbm{K}_1}{3\mathbbm{K}_0\sqrt{\lambda }}+\ldots \right).
\end{equation}
The constants $\mathbbm{K}_1$, $\mathbbm{K}_0$ are defined in (\ref{KK's}). Their ratio is just a number, to compute which we need to know $\hat{r}_{1a}(\xi )$. As we shall see, the expansion (\ref{expansioninKs}) is insufficient for this purpose.

For the second moment of the density we get:
\begin{equation}\label{x^2}
 \left\llangle x^2\right\rrangle=\frac{1}{6M^2}\int_{-\mu }^{\mu }dx\,r(x)\left(\mu ^4-x^4\right)=
 \frac{\mathbbm{K}_0\mu ^2}{4}+\frac{31\mathbbm{K}_1M\mu }{48}+\ldots ,
 \end{equation}
from which we can find the vacuum susceptibility:
\begin{equation}
 \chi =\frac{\left\llangle x^2\right\rrangle}{\left\llangle 1\right\rrangle}
 =\frac{\mu ^2}{4}\left(1+\frac{4\mathbbm{K}_1M}{3\mathbbm{K}_0\mu }+\ldots \right)
\end{equation}
The leading order is just the Gaussian average in the matrix model. Interestingly, the first correction to this result cancels when expressed in terms of the coupling constant:
\begin{equation}
 \chi =\frac{\lambda M^2}{16\pi ^2}\left(1+\frac{0}{\sqrt{\lambda }}+\ldots \right).
\end{equation}

The next term in the expansion (\ref{expansioninKs}) is of relative order $M/\mu \sim 1/\sqrt{\lambda }$, in accord with string-theory expectations, because $1/\sqrt{\lambda }$ is the natural parameter of the $\alpha '$ perturbation theory on the string worldsheet. The first correction to the effective string tension in (\ref{effstringT}) is $O(1/\sqrt{\lambda })$ in line with these expectations.
But the density of eigenvalues itself, it turns out, receives more subtle corrections at relative order $\sqrt{M/\mu }\sim \lambda ^{-1/4}$ which are hence more important than stringy $1/\sqrt{\lambda }$ corrections.

One may notice that (\ref{expansioninKs}) does not actually satisfy the correct boundary conditions (\ref{rasympt}).
This happens because
the expansion parameter in (\ref{expansioninKs}) is  not really $M/\mu $, but rather $M/\sqrt{\mu ^2-x^2}$. As soon as $x$ approaches one of the endpoints of the interval, the expansion breaks down and the simple approximation (\ref{expansioninKs}) then does not hold any more. To find the correct endpoint behavior of the master function, the integral equation (\ref{r-intrep}) has to be re-analyzed in the endpoint region, the importance of which for critical behavior has been recognized in \cite{Chen:2014vka}. To this end, we introduce a dimensionless function
\begin{equation}\label{what-is-R}
 R(\xi )=Mr(-\mu +M\xi ),
\end{equation}
defined in the endpoint region, that is for $\xi \sim 1$. As follows from (\ref{rasympt}),
\begin{equation}
 R(\xi )\rightarrow \frac{1}{\sqrt{\xi }}\qquad \left(\xi \rightarrow 0\right).
\end{equation}
Matching to the bulk solution (\ref{expansioninKs}) requires
\begin{equation}\label{infinity-R}
 R(\xi )\rightarrow \frac{\,{\rm const}\,}{\xi ^{\frac{3}{2}}}\qquad \left(\xi \rightarrow \infty \right).
\end{equation}

The main contribution to the function $\hat{r}_{1a}(\xi )$, as can be seen from the definition (\ref{xi-proj}), comes from the endpoint region. The boundary condition (\ref{infinity-R}) allows us to extend the summation range in the $\hat{\hphantom{r}}$-projection to infinity, leading to:
\begin{equation}
 \hat{r}_{1a}(\xi )\simeq \frac{2}{M}\sum_{l=0}^{\infty }R(l+\xi ).
\end{equation}
Substituting this result into (\ref{r-intrep}) and taking $x=-\mu +\eta M$, we get a self-consistent equation for $R(\eta )$:
\begin{equation}\label{aux-inteq}
 R(\eta )=\frac{1}{\pi \sqrt{\eta }}\int_{0}^{1}d\xi \,\,\frac{\sqrt{\xi }}{\eta +\xi }\,\sum_{l=1}^{\infty }R(l-\xi ).
\end{equation}
The solution to this equation with the prescribed boundary conditions is
\begin{equation}\label{near-b}
 R(\xi )=\frac{1}{\sqrt{\xi } }-\frac{1}{\sqrt{\xi +1}}\,.
\end{equation}
Then
\begin{equation}
 \sum_{l=1}^{\infty }R(l-\xi )=\frac{1}{\sqrt{1-\xi }}\,,
\end{equation}
and it is straightforward to check that eq.~(\ref{aux-inteq}) is indeed satisfied. 

For the functions $\hat{r}_{1a}(\xi )$ we get:
\begin{equation}
 \hat{r}_{1a}(\xi )\equiv \frac{2}{M}\sum_{l=0}^{\infty }R(l+\xi )=\frac{2}{M\sqrt{\xi }}\,,
\end{equation}
At this order there is no difference between $\hat{r}_{11}$ and $\hat{r}_{12}$.
Having $\hat{r}_{1a}(\xi )$, we can now calculate the constants $\mathbbm{K}_n$ defined in (\ref{KK's}). For the first two appearing in (\ref{expansioninKs}) we get:
\begin{equation}
 \mathbbm{K}_0=\pi \sqrt{\frac{2\mu ^3}{M^3}}
 \qquad 
  \mathbbm{K}_1=\frac{3\pi }{4}\,\sqrt{\frac{2\mu ^3}{M^3}}.
\end{equation}
In particular, from (\ref{effstringT}) we can calculate the first $1/\sqrt{\lambda }$ correction to the effective string tension:
\begin{equation}
 \mu =\frac{\sqrt{\lambda }M}{2\pi }\left(1-\frac{\pi }{\sqrt{\lambda }}+\ldots \right),
\end{equation}
reproducing the result of \cite{Chen:2014vka}.

At large $\xi $, the function (\ref{near-b}) behaves as expected from (\ref{infinity-R}) thus matching with the bulk solution (\ref{expansioninKs}), but at smaller $\xi $ the exact master function strongly deviates from the bulk approximation. The near-edge modification of the master function feeds back into the density. Computing the contribution from the boundaries in the projection  (\ref{rho-many-r}), (\ref{projection}) we get:
\begin{equation}
 \delta \rho (x)=\left.\frac{1}{C\mu M}\left[\left(\mu -x\right)R_1(\xi )+\left(\mu +x\right)R_1(\Delta -\xi +\theta (\xi -\Delta ))\right]\right|_{\xi =\left\{\frac{x+\mu }{M}\right\}},
 \end{equation}
 where
\begin{equation}\label{R1}
 R_1(\xi )=\sum_{l=1}^{\infty }\left(lR(l-1+\xi )-\frac{1}{2\sqrt{l}}\right)+\frac{1}{2}\,\zeta \left(\frac{1}{2}\right)=\zeta \left(\frac{1}{2}\,,\xi \right).
\end{equation}
Taking $C$ from (\ref{constant_C}), and combining this result with the leading order (\ref{rho-Wigner}), we get:
\begin{eqnarray}
 \rho (x)&\simeq&\frac{2}{\pi \mu ^2}\,\sqrt{\mu ^2-x^2}
\nonumber \\ &&
 +\frac{1}{\pi }\,\sqrt{\frac{M}{2\mu ^5}}
 \left[\left(\mu -x\right)\zeta \left(\frac{1}{2}\,,\xi \right)
 +\left(\mu +x\right)\zeta \left(\frac{1}{2}\,,\Delta -\xi+\theta (\xi -\Delta ) \right)\right].
 \label{two-oreders}
\end{eqnarray}
The second line is a correction of relative order $O(\sqrt{M/\mu })\sim O(\lambda ^{-1/4})$.  One may wonder if it were legitimate to keep the $O(M/\mu )$ terms in the normalization constant and the average in (\ref{various-averages}), (\ref{x^2}) while neglecting potentially more important boundary contributions. The leading boundary contributions to the normalization, the coupling and the second moment of the density, however, cancel as demonstrated in the appendix~\ref{appendix}. It would be interesting to understand at which order non-integer powers in $1/\sqrt{\lambda }$ show up in $\mu (\lambda )$.

The second term in (\ref{two-oreders}) describes the comb-like structure of resonances on top of the enveloping Wigner distribution. In fact, just these two terms give a very good approximation to the exact density for rather moderate $n$, for instance in the plot on fig.~\ref{wigner} the difference would be invisible as the error is smaller than the line thickness. The shape of each resonance is described by the zeta-function, in accord with the analysis of the edge behavior \cite{Chen:2014vka}. In the bulk, the zeta functions associated with left and right edges are added with coefficients $(\mu \pm x)$, which become small near the endpoints, and thus the left (respectively, right) resonances become increasingly  more pronounced towards the left (right) edge of the eigenvalue distribution, cf.~figs.~\ref{fig-n2}, \ref{fig-last} and \ref{wigner}.

Actually, the approximations used to arrive at the result (\ref{two-oreders}) break down near the endpoints \cite{Chen:2014vka}, where instead we have from (\ref{rho-many-r}), (\ref{xi-proj}):
\begin{eqnarray}
 MC\rho (x)&=&2\sum_{l=1}^{m-1}lR(l-1+\xi )+2m\sum_{l=m}^{\infty }R(l-1+\xi )
\nonumber \\
 &&-\frac{2m}{n}\sum_{l=1}^{\infty }lR(l-1+\xi )
 +\frac{2m}{n}\sum_{l=1}^{\infty }lR(l-1+\Delta -\xi +\theta (\xi -\Delta )),
\end{eqnarray}
 where, as before, 
\begin{equation}
 m=\left[\frac{x+\mu }{M}\right]+1,\qquad \xi =\left\{\frac{x+\mu }{M}\right\}.
\end{equation}
Using the explicit solution for $R(\xi )$, we get:
\begin{equation}
 \pi \sqrt{\frac{\mu ^3}{2M}}\rho (x)
 =\sum_{l=0}^{m}\frac{1}{\sqrt{\xi +l}}+\frac{m}{n}\left(
  \zeta \left(\frac{1}{2}\,,\Delta -\xi +\theta (\xi -\Delta )\right)-
 \zeta \left(\frac{1}{2}\,,\xi \right)
 \right).
\end{equation}
The first term reproduces the result of \cite{Chen:2014vka}. The second term is a $\lambda ^{-1/4}$ correction that contains a contribution from the opposite boundary, so that the density has singularities on both sides of the $a$-intervals, albeit one of them is parametrically stronger than the other.

\section{Conclusions}

The $\mathcal{N}=2^*$ SYM theory has a rich phase structure, which can be studied with the help of supersymmetric localization.  Although  at present the localization matrix model cannot be solved exactly in the strong-coupling phase, it is possible to get a fairly detailed picture of the eigenvalue density, which has rather irregular shape.  The critical behavior is associated with the mutations in the singularity structure of the density.

The most interesting open question is  how to describe this complex behavior  on the string side of the holographic duality. Since the majority of the phase transitions happen at strong coupling, one may hope to address this question without going far into quantum regime of the dual string theory.
In principle, the eigenvalue density can be probed rather directly in holography, by placing a D-brane in the dual geometry \cite{Buchel:2000cn,Evans:2000ct}. However, a classical D-brane seems to be sensitive to the averaged density and apparently cannot resolve the rigged and rather singular structure on the small scales \cite{Buchel:2000cn,Buchel:2013id}. The string description of the phase transitions observed in the matrix model thus remains an open problem, but since similar transitions appear quite generally in localization matrix models \cite{Russo:2013qaa,Russo:2013kea,Barranco:2014tla,Anderson:2014hxa,Russo:2014bda,Minahan:2014hwa,Marmiroli:2014ssa} one may hope to find equally universal holographic description.

\section*{Acknowledgements}

We would like to thank L.~Anderson for discussions and to L.~Anderson, X.~Chen-Lin and J.~Russo for comments on the manuscript.
This work was supported by the Marie
Curie network GATIS of the European Union's FP7 Programme under REA Grant
Agreement No 317089, by the ERC advanced grant No 341222
and by the Swedish Research Council (VR) grant
2013-4329.

\appendix

\section{Boundary contribution to averages}\label{appendix}

Here we show that the boundary contributions to $\left\llangle 1\right\rrangle$, $\left\llangle F(x)\right\rrangle$ and $\left\llangle x^2\right\rrangle$ cancel  at strong coupling to the leading order in $M/\mu $. 

For  $\left\llangle 1\right\rrangle$  we have from (\ref{compact-average}), (\ref{onehat}) and (\ref{what-is-R}):
\begin{equation}
 \delta \left\llangle 1\right\rrangle\simeq 2n\int_{0}^{1}d\xi \,\sum_{m=1}^{\infty }mR(m-1+\xi ).
\end{equation}
Regularization of the divergent sum over $m$ is dictated by matching to the bulk contribution. Careful analysis shows that the correct prescription is defined by (\ref{R1}), and then
\begin{equation}
 \delta \left\llangle 1\right\rrangle=2n\int_{0}^{1}d\xi \,R_1(\xi )=2n\int_{0}^{1}d\xi \,\zeta \left(\frac{1}{2}\,,\xi \right)=0.
\end{equation}
Similarly, from (\ref{xwithahat}) we find: 
\begin{equation}
 \delta \left\llangle x^2\right\rrangle\simeq \frac{M^2n^3}{18}\int_{0}^{1}d\xi \,R_1(\xi )=0.
\end{equation}

For the leading contribution to $\left\llangle F(x)\right\rrangle$ we get from (\ref{Fhat}):
\begin{equation}
 \delta \left\llangle F(x)\right\rrangle\simeq 2\int_{0}^{1}d\xi \,\sum_{m=1}^{\infty }R(m-1+\xi )\ln\frac{1-\xi }{\xi +m-1}\,.
\end{equation}
The sum now converges. Taking $R(\xi )$ from (\ref{near-b}), and doing the integral first, we get a telescoping sum, which leaves no boundary terms\footnote{A small subtlety here is that each term in the brackets is not formally defined at $m=1$ because of the logarithmic divergences. But the first term is multiplied by zero -- and should be set to zero -- while the divergences in the last two terms mutually cancel. The summand at $m=1$ should be understood as the limit of the expression in square brackets continued to non-integer $m$.}:
\begin{eqnarray}
 \delta \left\llangle F(x)\right\rrangle&=&4\sum_{m=1}^{\infty }\left[
 \sqrt{m-1}\ln(m-1)+\sqrt{m}\ln\frac{4m\left(\sqrt{m}-\sqrt{m-1}\right)}{\left(m-1\right)\left(\sqrt{m}+\sqrt{m-1}\right)}
 \right.\nonumber \\
&&\left.\vphantom{\ln\frac{4m\left(\sqrt{m}-
\sqrt{m-1}\right)}{\left(m-1\right)\left(\sqrt{m}+\sqrt{m-1}\right)}}
+\ln\frac{\sqrt{m}-1}{\sqrt{m}+1}
 -\left(m\rightarrow m+1\right)
 \right]=0,
\end{eqnarray}
so this average is also zero. In principle, to check the validity of (\ref{effstringT}), we need to calculate the next term in  $\left\llangle F(x)\right\rrangle$, but for that we need to know the boundary function $R(\xi )$ to the next-to-leading order accuracy, and this is beyond the scope of the present paper. 

\bibliographystyle{nb}

\begin{thebibliography}{10}
\ifx\href\asklfhas\newcommand{\href}[2]{#2}\fi
\raggedright
\small
\parskip 0pt

\bibitem{Pestun:2007rz}
V.~Pestun,
\textit{``{Localization of gauge theory on a four-sphere and supersymmetric
  Wilson loops}''},
\textsf{Commun.Math.Phys.~313,~71~(2012)},
\href{http://arXiv.org/abs/0712.2824}{\texttt{0712.2824}}.
%
\bibitem{Witten:1988ze}
E.~Witten,
\textit{``{Topological Quantum Field Theory}''},
\textsf{Commun.Math.Phys.~117,~353~(1988)}.
%
\bibitem{Erickson:2000af}
J.~K.~Erickson, G.~W.~Semenoff and K.~Zarembo,
\textit{``{Wilson loops in N = 4 supersymmetric Yang-Mills theory}''},
\textsf{Nucl.~Phys.~B582,~155~(2000)},
\href{http://arXiv.org/abs/hep-th/0003055}{\texttt{hep-th/0003055}}.
%
\bibitem{Drukker:2000rr}
N.~Drukker and D.~J.~Gross,
\textit{``{An exact prediction of N = 4 SUSYM theory for string theory}''},
\textsf{J.~Math.~Phys.~42,~2896~(2001)},
\href{http://arXiv.org/abs/hep-th/0010274}{\texttt{hep-th/0010274}}.
%
\bibitem{Russo:2013qaa}
J.~G.~Russo and K.~Zarembo,
\textit{``{Evidence for Large-N Phase Transitions in N=2* Theory}''},
\textsf{JHEP~1304,~065~(2013)},
\href{http://arXiv.org/abs/1302.6968}{\texttt{1302.6968}}.
%
\bibitem{Russo:2013kea}
J.~Russo and K.~Zarembo,
\textit{``{Massive N=2 Gauge Theories at Large N}''},
\textsf{JHEP~1311,~130~(2013)},
\href{http://arXiv.org/abs/1309.1004}{\texttt{1309.1004}}.
%
\bibitem{Buchel:2013id}
A.~Buchel, J.~G.~Russo and K.~Zarembo,
\textit{``{Rigorous Test of Non-conformal Holography: Wilson Loops in N=2*
  Theory}''},
\textsf{JHEP~1303,~062~(2013)},
\href{http://arXiv.org/abs/1301.1597}{\texttt{1301.1597}}.
%
\bibitem{Chen:2014vka}
X.~Chen-Lin, J.~Gordon and K.~Zarembo,
\textit{``{N=2* Super-Yang-Mills Theory at Strong Coupling}''},
\href{http://arXiv.org/abs/1408.6040}{\texttt{1408.6040}}.
%
\bibitem{Pilch:2000ue}
K.~Pilch and N.~P.~Warner,
\textit{``{N=2 supersymmetric RG flows and the IIB dilaton}''},
\textsf{Nucl.Phys.~B594,~209~(2001)},
\href{http://arXiv.org/abs/hep-th/0004063}{\texttt{hep-th/0004063}}.
%
\bibitem{Bobev:2013cja}
N.~Bobev, H.~Elvang, D.~Z.~Freedman and S.~S.~Pufu,
\textit{``{Holography for $N = 2^*$ on $S^4$}''},
\textsf{JHEP~1407,~001~(2014)},
\href{http://arXiv.org/abs/1311.1508}{\texttt{1311.1508}}.
%
\bibitem{Russo:2013sba}
J.~Russo and K.~Zarembo,
\textit{``{Localization at Large N}''},
\href{http://arXiv.org/abs/1312.1214}{\texttt{1312.1214}}.
%
\bibitem{Gross:1980he}
D.~Gross and E.~Witten,
\textit{``{Possible Third Order Phase Transition in the Large N Lattice Gauge
  Theory}''},
\textsf{Phys.Rev.~D21,~446~(1980)}.
%
\bibitem{Wadia:2012fr}
S.~R.~Wadia,
\textit{``{A Study of U(N) Lattice Gauge Theory in 2-dimensions}''},
\href{http://arXiv.org/abs/1212.2906}{\texttt{1212.2906}}.
%
\bibitem{Hoppe}
J.~Hoppe,
\textit{``{as cited in \cite{Kazakov:1998ji}}''}.
%
\bibitem{Kazakov:1998ji}
V.~A.~Kazakov, I.~K.~Kostov and N.~A.~Nekrasov,
\textit{``{D-particles, matrix integrals and KP hierarchy}''},
\textsf{Nucl.~Phys.~B557,~413~(1999)},
\href{http://arXiv.org/abs/hep-th/9810035}{\texttt{hep-th/9810035}}.
%
\bibitem{Kapustin:2009kz}
A.~Kapustin, B.~Willett and I.~Yaakov,
\textit{``{Exact Results for Wilson Loops in Superconformal Chern-Simons
  Theories with Matter}''},
\textsf{JHEP~1003,~089~(2010)},
\href{http://arXiv.org/abs/0909.4559}{\texttt{0909.4559}}.
%
\bibitem{Anderson:2014hxa}
L.~Anderson and K.~Zarembo,
\textit{``{Quantum Phase Transitions in Mass-Deformed ABJM Matrix Model}''},
\textsf{JHEP~1409,~021~(2014)},
\href{http://arXiv.org/abs/1406.3366}{\texttt{1406.3366}}.
%
\bibitem{Russo:2012kj}
J.~G.~Russo,
\textit{``{A Note on perturbation series in supersymmetric gauge theories}''},
\textsf{JHEP~1206,~038~(2012)},
\href{http://arXiv.org/abs/1203.5061}{\texttt{1203.5061}}.
%
\bibitem{Huang:2014pda}
X.~Huang and Y.~Zhou,
\textit{``{N = 4 Super-Yang-Mills on Conic Space as Hologram of STU Topological
  Black Hole}''},
\href{http://arXiv.org/abs/1408.3393}{\texttt{1408.3393}}.
%
\bibitem{Crossley:2014oea}
M.~Crossley, E.~Dyer and J.~Sonner,
\textit{``{Super-R\'enyi Entropy and Wilson Loops for N=4 SYM and their Gravity
  Duals}''},
\href{http://arXiv.org/abs/1409.0542}{\texttt{1409.0542}}.
%
\bibitem{Marmiroli:2014ssa}
D.~Marmiroli,
\textit{``{Phase structure of $\mathcal{N}=2^*$ SYM on ellipsoids}''},
\href{http://arXiv.org/abs/1410.4715}{\texttt{1410.4715}}.
%
\bibitem{Brezin:1977sv}
E.~Brezin, C.~Itzykson, G.~Parisi and J.~B.~Zuber,
\textit{``{Planar Diagrams}''},
\textsf{Commun.~Math.~Phys.~59,~35~(1978)}.
%
\bibitem{Russo:2012ay}
J.~Russo and K.~Zarembo,
\textit{``{Large N Limit of N=2 SU(N) Gauge Theories from Localization}''},
\textsf{JHEP~1210,~082~(2012)},
\href{http://arXiv.org/abs/1207.3806}{\texttt{1207.3806}}.
%
\bibitem{Gakhov}
F.~Gakhov,
\textit{``{Boundary value problems}''},
Dover Publications (1990).
%
\bibitem{Buchel:2000cn}
A.~Buchel, A.~W.~Peet and J.~Polchinski,
\textit{``{Gauge dual and noncommutative extension of an N=2 supergravity
  solution}''},
\textsf{Phys.Rev.~D63,~044009~(2001)},
\href{http://arXiv.org/abs/hep-th/0008076}{\texttt{hep-th/0008076}}.
%
\bibitem{Billo:2014bja}
M.~Billo, M.~Frau, F.~Fucito, A.~Lerda, J.~Morales et~al.,
\textit{``{Modular anomaly equations in N=2* theories and their large-N
  limit}''},
\href{http://arXiv.org/abs/1406.7255}{\texttt{1406.7255}}.
%
\bibitem{Evans:2000ct}
N.~J.~Evans, C.~V.~Johnson and M.~Petrini,
\textit{``{The Enhancon and N=2 gauge theory: Gravity RG flows}''},
\textsf{JHEP~0010,~022~(2000)},
\href{http://arXiv.org/abs/hep-th/0008081}{\texttt{hep-th/0008081}}.
%
\bibitem{Barranco:2014tla}
A.~Barranco and J.~G.~Russo,
\textit{``{Large N phase transitions in supersymmetric Chern-Simons theory with
  massive matter}''},
\textsf{JHEP~1403,~012~(2014)},
\href{http://arXiv.org/abs/1401.3672}{\texttt{1401.3672}}.
%
\bibitem{Russo:2014bda}
J.~G.~Russo, G.~A.~Silva and M.~Tierz,
\textit{``{Supersymmetric $U(N)$ Chern-Simons-matter theory and phase
  transitions}''},
\href{http://arXiv.org/abs/1407.4794}{\texttt{1407.4794}}.
%
\bibitem{Minahan:2014hwa}
J.~A.~Minahan and A.~Nedelin,
\textit{``{Phases of planar 5-dimensional supersymmetric Chern-Simons
  theory}''},
\href{http://arXiv.org/abs/1408.2767}{\texttt{1408.2767}}.
%
\end{thebibliography}

\end{document}